# Van der Waals pressure and its effect on trapped interlayer molecules

K. S. Vasu[1], E. Prestat[2], J. Abraham[1], J. Dix[3], R. J. Kashtiban[4], J. Beheshtian[5], J. Sloan[4], P. Carbone[3], M. Neek-Amal[5], S. J. Haigh[2], A. K. Geim[1] and R. R. Nair[1]

[1]School of Physics and Astronomy, University of Manchester, Manchester M13 9PL, UK.

[2]School of Materials, University of Manchester, Manchester M13 9PL, UK.

[3]School of Chemical Engineering and Analytical Science, University of Manchester, Manchester M13 9PL, UK.

[4]Department of Physics, University of Warwick, Coventry CV4 7AL, UK.

[5]Department of Physics, University of Antwerpen, Groenenborgerlaan 171, B-2020 Antwerpen, Belgium.

**Van der Waals assembly of two-dimensional (2D) crystals continue attract intense interest[1-5] due to the prospect of designing novel materials with on-demand properties. One of the unique features of this technology is the possibility of trapping molecules or compounds between 2D crystals[5,6]. The trapped molecules are predicted to experience pressures as high as 1 GPa[6,7]. Here we report measurements of this interfacial pressure by capturing pressure-sensitive molecules and studying their structural and conformational changes. Pressures of 1.2 ± 0.3 GPa are found using Raman spectrometry for molecular layers of one nanometer in thickness. We further show that this pressure can induce chemical reactions and several trapped salts or compounds are found to react with water at room temperature, leading to 2D crystals of the corresponding oxides. This pressure and its effect should be taken into account in studies of van der Waals heterostructures and can also be exploited to modify materials confined at the atomic interfaces.**

Van der Waals (vdW) interactions play a critical role in numerous phenomena and applications such as catalysis, adhesion, lubrication, nanofluidics and fabrication of novel vdW materials[1-10]. In the latter case, two-dimensional (2D) crystals assembled layer by layer are kept together only by vdW forces[1,11]. Various heterostructures composed of graphene, BN, $MoS_2$, etc. have been intensively investigated and already shown potential for a wide range of electronic and optical applications[3,4]. During the fabrication of such heterostructures, molecules become trapped between 2D crystals, which results in enclosures of nanometer height[5]. Such nano-enclosures were also fabricated intentionally, mostly using graphene, to serve as hydrothermal reactors[12,13], visualize chemical processes[14,15] and for electron and atomic force microscopy of biological samples and adlayers[6,16,17]. Hydrothermal reactions reported inside the graphene nanobubbles were limited to the pressure created at high temperatures where trapped solvents and molecules reach their critical or super critical phase to initiate the chemical reactions similar to the conventional hydrothermal anvil cell[13]. When nano-enclosures are created at room temperature, a pressure $P$ builds up, and it can be extremely high due to an exceptionally high stiffness of graphene and the fact that vdW forces become extremely strong on a sub-nm scale. The value of $P$ is determined by a balance between the resulting hydrostatic pressure, the deformation energy of 2D crystals and a gain in the adhesion energy[6,18]. The deformation energy favors enclosures with a flat top. In the first approximation, the pressure can be estimated as $\approx E_w/h$ where $E_w$ is the adhesion energy and $h$ is the height of the nano-enclosure[6,18]. For a trapped monolayer, this estimate yields $P$ of the order of 1 GPa, in agreement with further rigorous analysis[7,18,19]. However, such large vdW pressure has never been measured experimentally, nor has its effect on physical, structural and chemical properties of entrapped molecules been studied. Here we demonstrate the existence of the huge vdW pressure inside the nano-enclosures made from 2D crystals by studying changes in Raman spectra of pressure-sensitive molecules such as triphenyl amine (TPA) and boric acid (BA). We also report the effect of vdW pressure on the chemistry of materials inside nano-enclosures by investigating the chemical stability of trapped compounds [$MgCl_2$, $CsI$, $CuSO_4$ and $Ca(OH)_2$)] using both Raman spectroscopy and transmission electron microscopy (TEM).

Various 2D crystals have been used in our experiments to prepare nano-enclosures as described in detail in the supplementary information. For brevity, we focus below on enclosures made from an archetypal 2D material, graphene. Graphene-encapsulated (GE) molecule/salt structures were prepared by drop casting a small amount (2 μl) of a dilute ($\leq 0.1$ M) molecular or salt solution onto a graphene layer. A second (top) graphene crystal was then carefully placed on top using the standard transfer techniques[1,3] to trap the solution. We normally used few-layer graphene as a bottom layer to minimize its crumpling during the drop casting. The liquid between graphene crystals was allowed to dry overnight at room temperature and then further dried in a vacuum desiccator. As a result, the two graphene layers strongly attached to each other with a tiny amount of the solution being captured in between.

Fig. 1a shows an optical image of one of our GE-TPA samples. One can see many small bubbles (indicated with arrows in Fig. 1a) within seemingly featureless areas. However, atomic force microscopy (AFM) reveals that these areas are actually composed of flat nano-enclosures with a height of $\approx 1$ nm (region 1 in the inset of Fig. 1a) which are surrounded by numerous submicron bubbles of several nm in height (region 2). Such 1-nm-thick flat regions are found to be continuous, and their lateral size can be as large as several μm. A similar AFM landscape was found for GE-BA samples (see Supplementary Information).

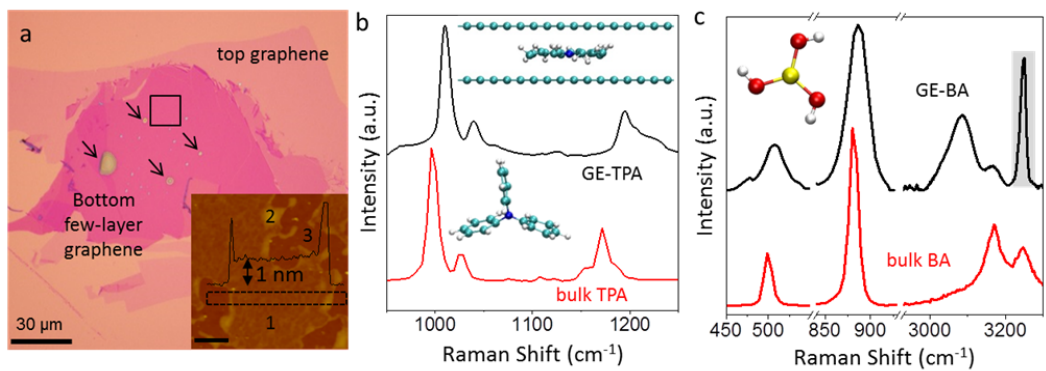

***Figure 1│Probing vdW pressure by Raman spectroscopy. a,*** *Optical micrograph of a graphene-encapsulated TPA on few-layer graphene placed on an oxidized Si wafer. Arrows indicates some of the small bubbles in the sample. Inset: AFM image of the region indicated by the square in (a). A 1-nm-thick flat region is marked as 1, one of many small bubbles marked as 2, and the region in which top and bottom graphene are attached as 3. Scale bar, 1 μm. Black curve: Height profile along the dotted rectangle.* ***b****, Typical Raman spectra from macroscopic bubbles and regions such as 1. The bottom inset shows the standard molecular structure of TPA. Top inset: modified structure of TPA between graphene sheets as found in MD simulations.* ***c****, Raman spectra of bulk BA and GE-BA. Inset: molecular structure of BA. The shaded area indicates the 2D' band of graphene.*

To determine the pressure inside this type of graphene nano-enclosure, we perform Raman spectroscopy. Fig. 1b compares spectra obtained from a bubble of several microns in size and from a flat 1-nm-thick region. The Raman spectrum for bubble agrees with the known spectrum of bulk TPA[20] where the major Raman bands are centered at ≈ 997 cm$^{-1}$ due to C–C stretching accompanied by weak C–N stretching, 1025 cm$^{-1}$ due to C–N stretching with associated C–C stretching, and 1172 cm$^{-1}$ due to combined C–N and C–C stretching. In stark contrast, the Raman spectrum of flat GE-TPA regions show all the bands being blueshifted with peaks appearing at ≈ 1011, 1040 and 1195 cm$^{-1}$, respectively. The widths of these bands are larger than those of bulk TPA. The observed changes in the Raman bands of captured TPA molecules are attributed to structural changes because of the confinement of TPA between graphene layers (Fig. 1b). Indeed, the propeller-shaped structure of TPA makes it highly sensitive to the uniaxial confinement, which can be translated into uniaxial pressure on the molecule. The observed Raman changes are in good agreement with those found for TPA at high pressures using conventional measurements in diamond anvil cells[21]. Comparison of our results with the latter measurements allows an estimate of *P* inside our flat graphene nano-enclosures as 1–1.5 GPa. To support this conclusion, we have also performed molecular dynamic (MD) simulations for TPA confined in a graphene capillary. They show that TPA molecules undergo a conformational change (from the propeller to planar structure; inset in Fig. 1b), if the separation between graphene sheets decreases below 1 nm (see Supplementary Information).

Similar experiments were performed with boric acid. Unlike TPA, BA molecules have a nearly planar structure already at ambient pressure. They are attracted to each other by hydrogen bonding and organize layered structures[22]. Fig. 1c shows Raman spectra obtained from bulk BA and from a 1-nm-thick enclosure with BA. The main bands for bulk BA are at 500 cm$^{-1}$ that corresponds to O–B–O angle deformation, 880 cm$^{-1}$ for B–O stretching, 3167 cm$^{-1}$ for symmetric O–H stretching and 3245 cm$^{-1}$ for antisymmetric O–H stretching[23]. For 1-nm-thick encapsulated BA, the two B–O modes exhibit broadening and a blueshift (509 and 885.5 cm$^{-1}$, respectively) whereas the O–H stretching modes show a redshift and a notable change in the shape of the peaks (Fig. 1c). Similar to the case of TPA, the observed changes in the Raman spectrum of flat GE-BA are consistent with previously reported hydrostatically pressurized BA[23]. Distinct from the other Raman bands of BA, the redshift for O–H stretching vibrations under pressure can be attributed to shortening of the hydrogen bond, which leads to lengthening and weakening of the parent O–H bond under compression[24]. The structure of BA is highly sensitive to pressure so that the O-B-O deformation and symmetric B-O stretching modes shift by 7.0 and 4.5 cm$^{-1}$ per GPa, respectively[23]. This allows us to estimate the pressure inside our 1-nm-thick graphene enclosures as $1.2 \pm 0.3$ GPa, in good agreement with the vdW pressure measured for TPA. Besides probing the vdW pressure, Raman spectroscopy also allowed us to estimate the strain developed in top layer graphene during the BA/TPA encapsulation by analyzing the shift in peak position of graphene G and 2D bands (see Supplementary Information).

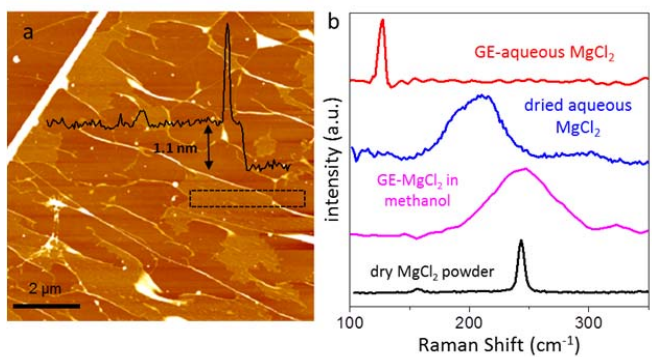

*Figure 2│Encapsulation-induced chemical reactions. a, AFM image obtained from graphene-encapsulated aqueous $MgCl_2$. Nanometer-thick flat regions are clearly seen being separated by wrinkles and occasional bubbles of triangular and arbitrary shapes. Black curve: Height profile along the dotted rectangle. b, Raman spectra measured from flat regions encapsulating $MgCl_2$ dissolved in water and methanol. For comparison, spectra for anhydrous and hydrated $MgCl_2$ are shown.*

The extremely high vdW pressure acting on substances captured between 2D crystals cannot be ignored and may modify physical and chemical properties of both trapped compounds and confining crystals experiencing the same pressure. As an example, we show that compounds such as $MgCl_2$, $CuSO_4$, and $Ca(OH)_2$ become reactive with water at room-temperature under the pressure inside nano-enclosures. Fig. 2a shows an AFM image of $MgCl_2$ solution trapped between graphene layers (also see Supplementary Information). Similar to the case of trapped TPA and BA, GE-MgCl$_2$ samples contain numerous flat regions with $h \approx 1$ nm. Fig. 2b plots typical Raman spectra acquired from such regions for 0.1 M solutions of MgCl$_2$ in water and methanol. As a reference, we also provide Raman spectra from an anhydrous MgCl$_2$ powder and a region where a large droplet of 0.1 M aqueous solution of MgCl$_2$ was allowed to dry up on top of graphene outside the nano-enclosures.

Anhydrous MgCl$_2$ exhibits two Raman bands, one at $\approx 240$ cm$^{-1}$ due to out-of-plane vibrations and another at $\approx 157$ cm$^{-1}$ assigned to in-plane phonons[25]. Also, the Raman spectrum obtained from the dried MgCl$_2$ solution shows typical signatures of hydrated MgCl$_2$[26] (see Supplementary Information) with a broad red-shifted peak at $\approx 200$ cm$^{-1}$. In contrast, the Raman spectrum of 0.1 M aqueous MgCl$_2$ inside the flat GE region does not show any of the expected features but instead a sharp peak appears at $125 \pm 3$ cm$^{-1}$. This new Raman band was universally observed for many flat encapsulated regions, reaching several microns in size. In addition to the 125 cm$^{-1}$ band, we also observe changes in the Raman spectrum of graphene covering MgCl$_2$ regions, which include the emergence of the D peak (see Supplementary Information). In comparison, nano-enclosures with MgCl$_2$ dissolved in methanol show only broadened Raman signature characteristic of anhydrous MgCl$_2$ and no sign of the 125 cm$^{-1}$ peak (Fig. 2b).

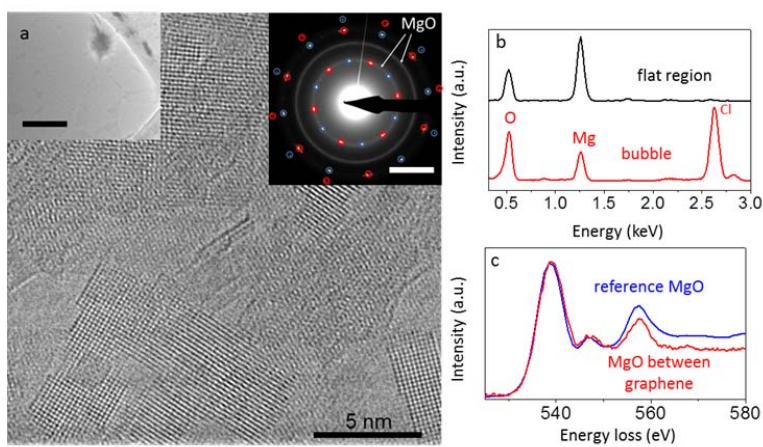

*Figure 3│TEM on graphene-encapsulated aqueous MgCl$_2$. **a**, High-resolution TEM image of MgO nanocrystal formed between two graphene layers. Left inset: Low-magnification TEM image showing flat encapsulated regions and a larger bubble. Scale bar, 500 nm. Right inset: Diffraction pattern taken from a 1 μm-diameter flat area shows two hexagonal patterns (red and blue circles), which come from the top and bottom graphene layers, and additional diffraction rings corresponding to the {200} and {220} planes of polycrystalline MgO. Scale bar, 5 nm$^{-1}$. **b**, EDX spectra from a flat encapsulated region and from a large bubble. **c**, Oxygen-K edge spectrum from nanocrystals such as in (a) compared with the spectrum of bulk MgO from the EELS database (https://eelsdb.eu/spectra/magnesium-oxide-2).*

Although the described spectroscopic changes indicate dramatic changes in the structure of the trapped aqueous $MgCl_2$, Raman spectroscopy is unable to reveal what kind of transformation takes place. To elucidate this, we have performed TEM analysis of constituents inside the nano-enclosures (see methods). Our low-magnification TEM images confirm that the samples contain flat enclosed areas with only a few small bubbles in between (inset of Fig. 3a), in agreement with our AFM imaging. Elemental analysis using energy dispersive X-ray (EDX) spectroscopy shows that the bubbles contain magnesium, oxygen and chlorine, as expected for aqueous $MgCl_2$, whereas the flat enclosures contain magnesium and oxygen but surprisingly with no signature of chlorine (Fig. 3b). High-resolution TEM also reveals that the flat regions are thin crystallites of one to few layers in thickness, which have a square lattice with lattice constant of 2.1 ± 0.05 Å (Fig. 3a). Electron diffraction and electron energy loss spectroscopy (EELS) analyses confirm that these nanocrystals are MgO, with the EELS oxygen K-edge from the enclosed material showing a good match to the reference spectrum for bulk MgO (see Fig. 3c). Furthermore, fast Fourier transform of high-resolution image of the thinnest MgO crystallites show the presence of a (110) lattice reflection, and our simulated electron diffraction patterns confirm that such (110) reflections are a clear signature of monolayer MgO (see Supplementary Information). All this unequivocally shows that the observed nanocrystals inside graphene nano-enclosures are MgO. This conclusion is also consistent with the observed 125 $cm^{-1}$ Raman band that was previously reported for MgO crystals[27,28].

We have performed similar experiments with aqueous solutions of other compounds including $CuSO_4$, CsI and $Ca(OH)_2$ (see Supplementary Information). Similar to the case of $MgCl_2$, our Raman and TEM analyses show that graphene-encapsulated $CuSO_4$ and $Ca(OH)_2$ are converted into CuO and CaO, respectively, whereas graphene-encapsulated CsI remains stable. This agrees with our density functional theory calculations performed for $MgCl_2$, $Mg(OH)_2$ and CsI trapped with water inside a hydrophobic cage (see Supplementary Information). They suggest the formation of a Mg–O bond in the case of $MgCl_2$ and $Mg(OH)_2$ and no changes for CsI. Although the calculations do not elucidate the underlying mechanism for the room-temperature reactivity of the salts with $H_2O$, it is reasonable to assume that the reactions are induced by the high vdW pressure. Indeed, the probability of ionization of water molecules strongly increases with increasing $P$, and changes in pressure from 0.1 MPa to 1 GPa are known to cause the ionization constant of water to change by two orders of magnitude[29], thus favouring the hydrolysis. We propose that the observed room-temperature reaction of nano-enclosed salts with water involves the pressure-assisted hydrolysis of salts into their hydroxides followed by decomposition to the corresponding oxides (see Supplementary Information). Conversion of bulk salts to corresponding hydroxides or oxides was only observed previously at high-temperatures and pressures[30] and we assume that the confinement and pressure has a significant influence on the feasibility of observed reaction at room temperature. The effect of confinement apparently brings the close proximity of metal ions to ionized water molecules to begin hydrolysis reaction and it is important to note that without the van der Waals confinement the pressure alone could not make the observed reaction feasible at room temperatue[31,32]. The reaction byproducts such as HCl and water are probably accumulated in the small bubbles and wrinkles observed in the flat encapsulated regions (see Supplementary Information) or evaporated through microscopic defects created in graphene during the reaction as the presence of graphene's D peak indicates (see Supplementary Information). The proposed mechanism is also consistent with the absence of changes for CsI. In this case, the salt of strong acid and strong base completely ionizes in water rather than undergoing hydrolysis[33]. In contrast, $MgCl_2$ and $CuSO_4$ are examples of salts of a weak base and strong acid that normally undergo hydrolysis (see Supplementary Information). The observed non-reactivity for $MgCl_2$ in methanol is naturally explained by the absence of water. Still, further experimental and theoretical studies are needed for a better understanding of the chemical transformations inside graphene nano-enclosures.

In conclusion, our work shows that huge pressures are exerted on materials trapped at interfaces during assembly of vdW heterostructure and that these pressures can lead to unexpected physical and chemical changes. This should be considered in analysis of the properties of vdW heterostructures, especially because the neighboring 2D crystals experience the same pressure. Our results can be exploited to induce novel physical phenomena in nano-enclosures and create new chemical compounds, which all points at many opportunities to explore the science of nanoconfined materials.

**Methods**

**Fabrication of graphene-encapsulated salt or molecule samples.** Single layer and few-layer graphene flakes obtained by mechanical exfoliation were used to fabricate the GE salt or molecule samples as described in detail in the supplementary information. Single layer graphene supported on poly(methyl methacrylate) (PMMA) layer prepared by wet transfer technique (see supplementary information) was used as a top layer to enclose the salt or molecule solution placed on another graphene or graphite flake exfoliated on $Si/SiO_2$ substrate. The encapsulation process was carried out using standard micromanipulation setup followed by drying of the sample and removal of PMMA in acetone wash. For TEM imaging and analysis, samples were prepared using the single layer graphene grown on copper via chemical vapor deposition (see Supplementary Information). We have used 0.001 M to 0.1 M solutions of the molecule or salt, and all reported results were similar even when the different concentrations of molecule or slat solutions were used. All sample solutions were aqueous, except triphenylamine (TPA), which is insoluble in water so methanol was used as the solvent.

**Atomic force microscopy (AFM) measurements.** AFM imaging of the encapsulated samples was performed using a Bruker Dimension FastScan AFM operating in peak force tapping mode.

**Raman measurements.** A micro Raman setup consisting of an optical microscope, and a spectrograph with 2,400 and 1,800 lines/mm grating (Reinshaw/Witec spectrometer) was used to obtain Raman spectra in a backscattering geometry using a 100× objective lens (NA = 0.70) under ambient conditions. An Ar-ion multiline tuning laser operating at wavelengths of 488 and 514 nm and a He–Ne laser operating at a wavelength of 633 nm were used as the excitation sources for the Raman measurements in this study. We chose a laser power well below 2 mW in order to avoid any laser-induced heating effects during the measurements. For the Raman experiments, thick graphite flakes were used as the bottom layer in the encapsulated samples to avoid the background Raman spectrum of the Si/SiO$_2$ substrate.

**TEM imaging.** Aberration-corrected TEM, energy-dispersive X-ray (EDX) spectroscopy and electron diffraction (ED) experiments were carried out using a JEOL ARM 200F TEM microscope operated at an operating voltage of 80 kV. The dose rate for high resolution transmission electron microscope (HRTEM) imaging was maintained at $10^4$ electrons per Å$^2$ per second. EDX spectroscopy was performed using the ARM's window-less Oxford Instruments X-MaxN 100TLE detector and TEM images were recorded using a Gatan Orius CCD camera.

An FEI Titan 80-200 ChemiSTEM equipped with probe-side aberration correction and an X-FEG electron source was used for the aberration-corrected scanning transmission electron microscope (STEM) imaging, electron energy loss spectroscopy (EELS) and ED analysis. STEM experiments were performed using an acceleration voltage of 200 kV, a beam current of 150 pA and a convergence angle of 21 mrad. EDX spectrum imaging was performed using the Titan's four detector Super-X spectrometer system, providing a solid angle of approximately 0.7 sr. EELS data was acquired using a GIF Quantum ER Spectrometer with an energy dispersion of 0.25 eV and a collection angle of 62 mrad, providing an effective energy resolution of 1.8 eV. In standard STEM imaging mode, the high angle annular dark field (HAADF) collection angles were 62– 142 mrad, while during spectrum imaging the HAADF collection angles were 60–190 mrad. Electron diffraction patterns were acquired with a beam current of 2 nA and with an illuminated area on the specimen of ≈ 1 μm$^2$.

# Supplementary Information

## Van der Waals pressure and its effect on trapped interlayer molecules

1. Fabrication of graphene-encapsulated salt or molecule samples

Graphene-encapsulated (GE) samples were prepared by using the wet transfer technique[1] to enclose salt or molecule solutions between two micromechanically exfoliated graphene flakes or between a graphene flake and a graphite flake as summarized in Fig. S1. Single- and few-layer graphene flakes were prepared on Si/SiO$_2$ substrates with an oxide thickness of 300 nm using the micromechanical exfoliation[2]. Optical microscopy was used to locate the graphene flakes of suitable size (50 μm × 50 μm or above) on Si/SiO$_2$ substrate (panel 1 in Fig. S1) and the thickness of these flakes was verified using Raman spectroscopy[3]. A thin layer (≈ 200 nm) of A3-950 poly(methyl methacrylate) (PMMA) resist was then spin coated on to a substrate containing the desired flake and subsequently heated at 100 °C for 10 min to evaporate the solvent from the resist (panel 2 in Fig.S1). An adhesive tape window was placed above the PMMA layer, ensuring that the desired graphene flake was at the center of the open area (panel 3 in Fig. S1). The tape and PMMA act as a support for the graphene flake allowing it to be lifted away from the Si/SiO$_2$ substrate. PMMA layer at the periphery of the tape window was then removed in order to separate the tape window area from the remainder of the PMMA layer. The entire sample was then placed in 3% KOH solution to etch away the SiO$_2$ layer and thereby separate the graphene flake from the substrate (panel 4 in Fig. S1). After this etching process, the tape window with the PMMA layer containing the graphene flake floated on the surface of the KOH solution due to the hydrophobic nature of PMMA, whereas the Si substrate stays at the bottom (panel 5 in Fig. S1). Subsequently, the tape window hosting the graphene-PMMA layer was rinsed with deionized water to remove any residual KOH solution. Using the micromanipulation setup[4], this graphene–PMMA layer was carefully placed on top of 2 μl of the molecule or salt solution casted onto the bottom few-layer graphene (FLG) or graphite flake. A single layer graphene (SLG) has also been tested as a bottom layer but was not ideal as it often ruptured during drop casting of the molecule or salt solution. After placing the single layer of graphene on top of the solution, most of the solution was spontaneously squeezed out by leaving only a very small amount in between the top and bottom graphene layers. The prepared samples were dried overnight at room-temperature in order to let the solvent to evaporate gradually and allow the top graphene layer to collapse onto the bottom graphene flake. These samples were then placed in vacuum (~1 mBar) for several hours to remove any residual solvent and water molecules. Finally, the tape window was detached, and the top PMMA layer was removed using an acetone wash.

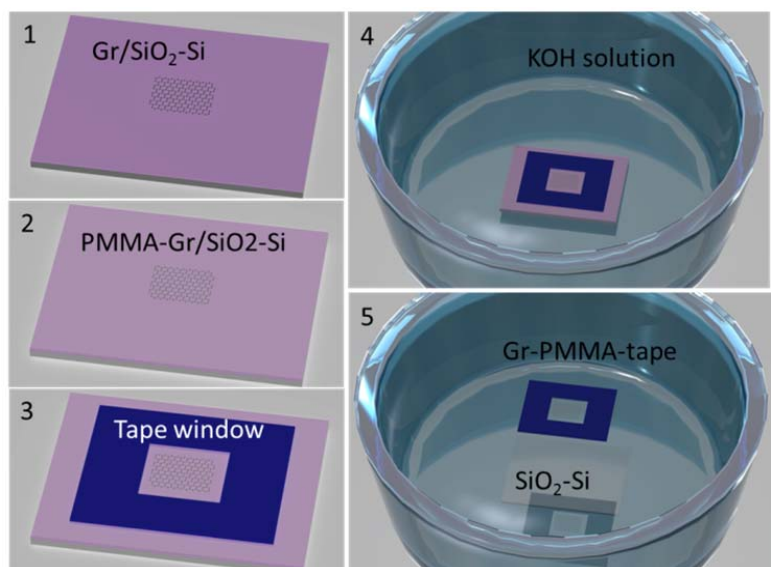

**Figure S1. Wet transfer of graphene.** Schematic illustration of the steps used for producing a top layer of graphene (Gr) needed for fabrication of an encapsulated sample. Panel numbers indicate the sequence of the steps in the process (1-5).

Graphene grown using chemical vapor deposition (CVD) on Cu foil, received from BGT Materials Limited (Manchester, UK), has been used to prepare samples suitable for transmission electron microscope (TEM) and scanning TEM high-resolution imaging and analysis. To detach the CVD grown graphene from the Cu foil (5 mm × 5 mm), the graphene was coated with PMMA and the metal was etched away using a 0.1M ammonium persulfate aqueous solution. The resulting graphene-PMMA film was then rinsed several times in water and transferred onto a gold mesh Quantifoil TEM support-grid. Removal of the PMMA layer was achieved by washing in acetone and isopropyl alcohol (IPA) baths. The graphene covered TEM grids produced were then dried in a critical point dryer and subsequently annealed in activated carbon at 250 °C to remove the hydrocarbon contamination[5]. To prepare the GE salt samples for TEM, salt solution was drop casted at the centre of a graphene-covered TEM grid followed by placing another graphene-covered TEM grid above the droplet. The prepared TEM samples were further dried as described above.

2. Atomic force microscopy (AFM) of graphene-encapsulated salt and molecule

Fig. S2a shows the AFM image of a GE boric acid (BA) sample on few-layer graphene showing 0.9 nm-thick, large, flat, BA-enclosed regions and small nano-bubbles. Similar to the case of TPA (see the main text), the AFM image shows that the BA-enclosed flat regions are continuous with dimensions reaching up to 10 μm$^2$ in area. Similar features were also observed in GE-$MgCl_2$ samples. However, unlike images of GE-TPA or BA samples, the AFM image (Fig. S2b) taken of a flat, encapsulated region of the GE-$MgCl_2$ showed small bubbles or wrinkle features. We believe that these are caused by gaseous phase by-products produced during the conversion of $MgCl_2$ to MgO (see main text).

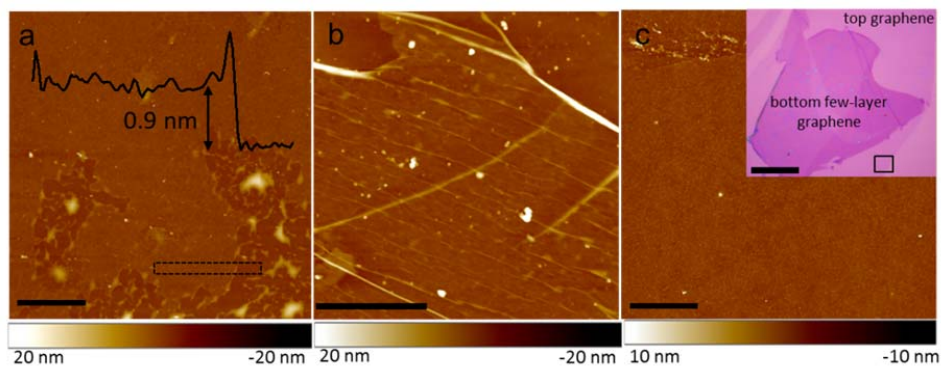

**Figure S2. AFM images taken for graphene-encapsulated salt and molecule on few-layer graphene.** (a) AFM image showing a 0.9-nm-thick, large, flat region in a GE-BA sample. Dark curve: height profile along the dotted rectangle. (b) AFM image showing wrinkle features within the flat 1-nm-thick $MgCl_2$-captured region in a GE-$MgCl_2$ sample. (c) AFM image in the same GE-$MgCl_2$ sample as shown in Fig. S2b showing absence of bubbles or flat encapsulated regions outside of the graphene-graphene enclosure (in an area where the top single layer of graphene lies on the Si/$SiO_2$ substrate). All scale bars are 1 µm. Inset shows the optical micrograph of the GE-$MgCl_2$ sample on few-layer graphene (scale bar 30 µm), where the square marked region indicates the area from which the AFM image was obtained.

We further investigated the possibility of capturing salt and other molecules between the graphene and Si/$SiO_2$ substrate. Fig. S2c shows the AFM image taken of a GE-$MgCl_2$ sample in a region where the $MgCl_2$ is expected to be enclosed between the top single-layer graphene and bottom $SiO_2$ substrate (indicated by a black square in the inset of Fig. S2c). The absence of either bubbles or flat captured regions in this area suggests that the $SiO_2$/graphene interface does not favour capturing the molecules, possibly due to the relatively high roughness of the Si/$SiO_2$ substrate; molecules can diffuse along the $SiO_2$/graphene interface.

3. Estimation of strain on the top layer graphene

Fig. S3 shows the Raman spectra of graphene (collectively from top and bottom layers) obtained from the three different regions of GE-BA sample including the collapsed graphene region, 1-nm-thick flat BA enclosed region and a nanobubble of height 60 nm and base 1.5 µm. G band in the Raman spectrum of collapsed region appears as a single mode due to merging of G band from both the top SLG and bottom FLG, whereas 2D band looks like a convolution because the peak position and intensity of 2D band from SLG and FLG are certainly distinctive. In contrast, Raman spectrum obtained from the 1-nm-thick flat enclosed region and nanobubble shows broadening/splitting of G and 2D bands due to the redshift in the SLG Raman modes caused by the isotropic strain developed in top SLG during the encapsulation. For example, in the case of nanobubble, G peak position is decreased from 1583 $cm^{-1}$ to 1542 $cm^{-1}$ while 2D peak has shifted to 2593 $cm^{-1}$ from 2683 $cm^{-1}$. Based on the reported Raman shift of G and 2D peak positions with respect to the strain[6], we have estimated the strain developed on top SLG at this bubble and flat 1-nm-thick enclosures as ~ 0.7% and 0.05-0.1% respectively.

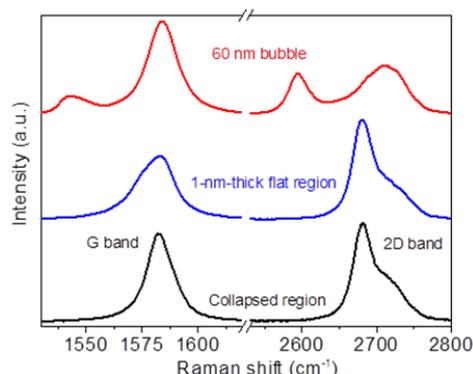

**Figure S3. Strain induced shift in the Raman spectra of graphene.** Raman spectrum (514 nm excitation) of graphene (collectively from top single layer and bottom few-layers) obtained from a collapsed region, 1-nm-thick flat enclosed region and nanobubble of height 60 nm in a GE-BA sample showing the strain induced shift in the top layer graphene.

4. Molecular dynamic simulation of TPA between graphene

The molecular dynamics (MD) simulations were conducted using the GROMACS 4.5.4 package[7]. MD simulations used a list cut-off and short-range coulomb cut-off of 1.3 nm. For the van der Waals (vdW) interactions, we have used a switching function to allow the potential to go to zero. The switching function starts at 0.9 nm and causes the vdW force and potential to go to zero at 1.0 nm. For all simulations, we use a 1 fs time step with the leapfrog integrator[8]. The initial setup for the simulations is shown in Fig. S4, where two parallel graphene sheets with a sheet size of 5.16 × 5.10 nm are fixed with an inter sheet separation of 1.3 nm and five TPA molecules are placed within the channel. The remaining volume of the channel was filled with methanol, and the entire channel is placed into contact with a methanol reservoir (8 nm × 8 nm × 8 nm). The graphene sheets are not allowed to bend in the direction perpendicular to the plane of the sheet, which fixes the graphene–TPA/methanol interfacial area. Edges of the channel are open so that the methanol in the reservoir can exchange with the methanol in the channel and reach thermodynamic equilibrium. These simulations were run in the isothermal–isobaric ensemble at a temperature and pressure of 298.15 K and 1 bar, respectively. This was run for 7 ns, with the Berendsen thermostat[9] and barostat[9]. Carbon–hydrogen bond lengths were maintained at 1.08 Å and 1.09 Å for TPA and methanol, respectively, using the LINCS constraint algorithm. The oxygen–hydrogen bond length for methanol was constrained to 0.945 Å using the LINCS[10] constraint algorithm. A Particle–Mesh Ewald (PME)[11,12] summation was used to consider the long-range electrostatics. All of the atomic parameters for the bonding and non-bonding interaction of methanol and TPA were taken from the OPLS–AA force field[13,14]. The carbon atoms of the graphene sheet had no partial charge but did have Lennard–Jones 12–6 interaction parameters[13,15] of $\sigma_c$= 0.355 nm and $\varepsilon_c$= 0.29288 kJ mol$^{-1}$. This is based on the OPLS–AA parameters for aromatic carbon species. The Lennard–Jones interaction parameters between different atoms were determined using the geometric mixing rules.

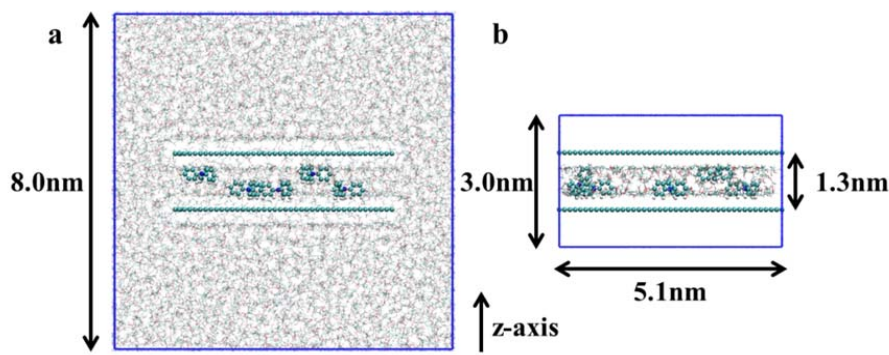

**Figure S4. Starting configurations of the molecular dynamics simulation.** The images show the starting configurations for the reservoir simulation (a) and the infinite channel simulation (b) for GE-TPA molecules in the presence of methanol. Solid blue lines mark the edge of the simulation cell. The dimensions of the box are 8 nm × 8 nm × 8 nm for **a** and 5.1 nm × 5.1 nm × 3.0 nm for **b**. Separation between graphene sheets in both images is 1.3 nm. The z-axis is defined as perpendicular to the graphene sheets. Color key: C – Cyan, H – White, N – Blue, O – Red.

After the equilibrium was reached, the reservoirs were removed, and all molecules in the channel were placed into an infinite slit pore in which the graphene sheets were 1.3 nm apart, as shown in Fig. S4b. The slit pore was seen as infinite due to the periodic boundary conditions applied in the directions parallel to the graphene sheets. The simulation was run in the canonical ensemble with constant box dimensions of 5.1 nm × 5.1 nm × 3.0 nm and at a constant temperature of 298.15 K. To maintain the temperature, a Nosé–Hoover[16,17] thermostat was used with a temperature coupling constant of 0.2 ps. The z-dimension was kept larger than the channel width in order to prevent periodic images from interacting with each other perpendicularly to the graphene sheets. These simulations had an initial energy minimization step to ensure that there were no high-energy overlaps at the start of the simulation. Subsequently, the sheets were pulled together at a rate of $0.0001\ nm\ ps^{-1}$, and the separation was changed from 1.3 nm to lower values. For this simulation, the PME summation is used in only 2 dimensions—those in the plane of the graphene sheet.

The orientations of the TPA molecules with respect to the graphene sheets were measured for different graphene separations as shown in Fig. S5. To monitor the change in the orientation of the TPA molecules during the simulations, we have looked at the angle between the vector normal to the graphene sheet and normal to the phenyl rings. This angle can be measured from either of the graphene sheets but has been restricted to always be from 0–90º, allowing the 'TPA angles' to be averaged over all of the molecules. For each point on Fig. S5 the TPA angle at a particular inter sheet separation is averaged over 100 snapshots taken at 100 fs time intervals. The error for each point was also calculated, but found to be smaller than the thickness of the line in the plot (Fig. S5) so has not been directly included. Around a channel width of 1.1–1.3 nm there is no favourable orientation of the phenyl rings with respect to the plane of the graphene sheets and there is a large variation in the average angle between the phenyl rings and graphene sheets. By compressing the channel width to 0.9–1.0 nm, the phenyl rings are forced to orient themselves perpendicular to the graphene sheet giving higher TPA angles ≈ 75º (see Fig. S5 inset). Under further compression, at a channel width of 0.7–0.8 nm, the phenyl rings lie at an angle of ∼ 45º angle to the graphene sheets, and further confinement leads to alignment of the TPA parallel to graphene sheets (see Fig. S5 inset).

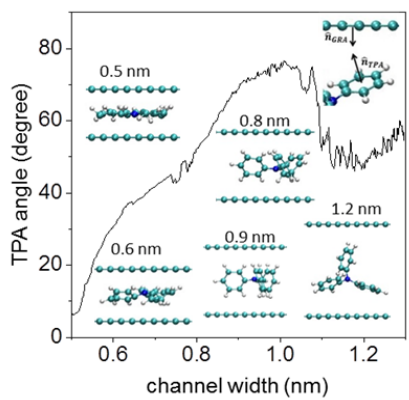

**Figure S5. Confinement-induced effects on TPA measured using MD simulations.** Graph shows TPA angle measured as the angle between the vector normal to the TPA phenyl rings and the vector normal to the graphene sheet. Top right inset shows an example of these two vector normals. This TPA angle is average over all phenyl rings for all of the TPA molecules in the simulation. All other insets show the structure of TPA inside graphene capillary under different confinement separations (0.5 – 1.2 nm).

From the infinite channel simulations it is possible to determine the pressure on confined molecules between the graphene sheets at different channel widths from the trace of the rank-2 stress sensor[18,19] (see Fig. S6). Simulations were further carried out by varying the strength of interaction between the graphene sheets and confined molecules to investigate the effect of interaction strength on the observed vdW pressure inside the nano-enclosures. This was done by changing the Lennard-Jones 12-6 parameters for the C atoms of the graphene sheets. Values of $\sigma_c$ = 0.321 nm and 0.355 nm were used with values of $\varepsilon_c$ = 0.05 kJ mol$^{-1}$, 0.29 kJ mol$^{-1}$ and 0.60 kJ mol$^{-1}$, giving 6 combinations of different parameters and this range of values will provide us a better insight on the interactions between confined molecules and graphitic surface. It has been shown that this range of parameters turn graphite surface from hydrophobic to hydrophilic as $\varepsilon_c$ varied from 0.05 kJ mol$^{-1}$ to 0.60 kJ mol$^{-1}$ with $\sigma_c$ = 0.321 nm[20]. The pressure on entrapped liquid molecules was then monitored as a function of the interlayer sheet distance. Since the graphene atoms were kept fixed at their positions, the forces acting on them are not included in the calculation of the pressure. As molecules are compressed into the channel, pressure inside the channel increases (see Fig. S6). The plot shows that for interlayer separation below 0.9 nm ca. the pressure starts to increase very rapidly reaching a value of 1 GPa depending on the value of $\sigma_c$. Indeed, in a given nano-enclosure the smaller graphene carbon atoms ($\sigma_c$ = 0.321 nm) reduce the pressure because the confined molecules have more volume to occupy. Increasing the value of $\varepsilon_c$ instead induces a modest increase in pressure since the carbon atoms are more attractive at long distances, but more repulsive at shorter distances (i.e. the potential energy increases more rapidly) and therefore at large confinement (below 0.8 nm), where the majority of the interaction with the graphene sheets are repulsive, the pressure increases. Considering that the experimental interlayer distance from the AFM measurements is an average value over an area of few nanometers, and the force field (OLPS) used in the simulation is parameterized to reproduce density of trapped molecules at atmospheric pressure but not at high pressure, our simulation results agree reasonably well with the experimental data.

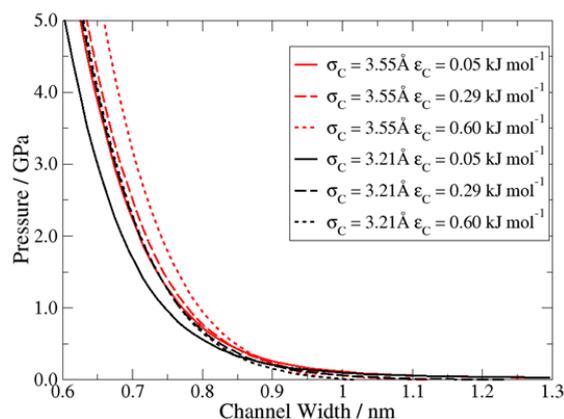

**Figure S6. Effects of graphene-molecule interaction on pressure.** Graph shows the pressure measured in simulation against the channel width for different graphene interaction parameters.

5. Raman spectrum of $MgCl_2$ upon hydration

$MgCl_2$ is known to be highly hygroscopic, causing the thin $MgCl_2$ films to absorb water molecules quickly from the atmosphere. The Raman spectrum (Fig. S7) obtained from a dried thin layer of $MgCl_2$ formed by drop casting the salt solution on $Si/SiO_2$ and vacuum drying has shown two Raman bands at ≈157 and 240 cm$^{-1}$ which are characteristic of $MgCl_2$[21]. We have also observed a less intense Raman bands around 3100-3600 cm$^{-1}$ correspond to the O–H bending modes. These modes are expected to arise from the residual water molecules left in the dry $MgCl_2$ thin layer; the very high hygroscopic nature of $MgCl_2$ prevents the water molecules from being evaporated completely. When the dried sample was later exposed to air the above mentioned Raman bands of $MgCl_2$ disappeared, and a new band at ≈ 200 cm$^{-1}$ with a full width half maximum (FWHM) of 60 cm$^{-1}$ appeared along with the high-intensity O–H Raman bands. These observed features are in good agreement with Raman spectra reported[21,22] for the hydrates of $MgCl_2$.

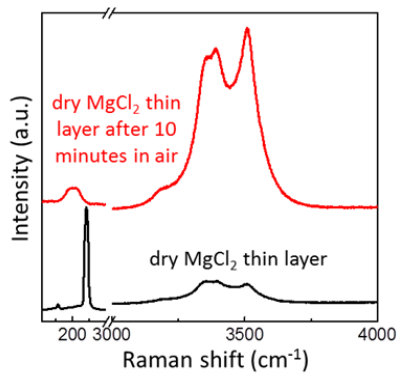

**Figure S7. Raman spectrum of hydrated $MgCl_2$.** Raman spectrum (488 nm excitation) obtained from a thin layer of $MgCl_2$ deposited on a $Si/SiO_2$ substrate by drop casting 0.1 M $MgCl_2$ solution immediately after vacuum drying and after 10 min of exposure to air.

## 6. Electron diffraction simulations

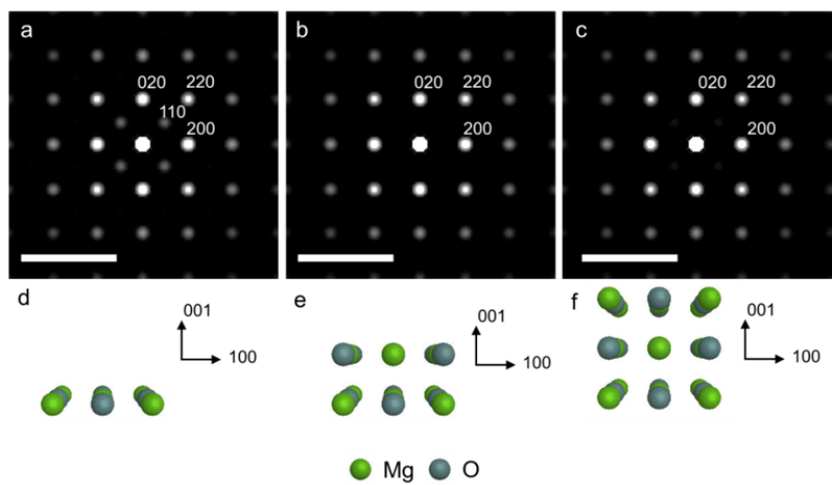

**Figure S8. Simulations of the MgO electron diffraction pattern:** (a) Simulated electron diffraction patterns from single-layer MgO (b) bilayer MgO and (c) trilayer MgO oriented along the [001] zone axis (scale bar 10 nm$^{-1}$). (d, e and f) Atomic models used for simulating single layer, bilayer and trilayer MgO, respectively.

Electron diffraction patterns for single, bilayer and trilayer MgO sheets were simulated using the JEMS software[23] using the kinematical approximation (Fig. S8a-c). The corresponding atomic models used for these simulations are shown in Fig. S8 (d-f). Because of the structure factor of the bulk MgO crystal (rock salt structure, space group $Fm\bar{3}m$, No. 225), the (110) lattice reflection is forbidden. However, these simulations show that for the single-layer case, the forbidden (110) reflection is present, whereas in the bilayer and trilayer cases, this reflection is absent and extremely weak, respectively. Our simulation also shows that the (110) reflections will be present for MgO crystals with an odd number of layers, although the intensity decreases rapidly with increasing sheet thickness.

Fig. S9a shows a HRTEM image obtained from the thinnest MgO crystals encapsulated between two graphene sheets and the corresponding fast Fourier transform (FFT) image showing the lattice spacing corresponding to the graphene sheets and to the MgO crystals. The presence of (110) MgO reflections suggests that the MgO crystals are single layers. We have not clearly observed the (110) forbidden reflections in the experimental diffraction pattern shown in Fig. 3 in the main text. This may be due to the weak intensity of the (110) reflection, which is approximately 10 times lower than the (200) reflections for monolayer materials (see Fig. S8a). Additionally, the experimental diffraction patterns are acquired over a much larger area (~ 1 µm²), where the thickness is non-uniform and which predominantly contains few layers of MgO. Consequently, the diffraction signal from the monolayer MgO is expected to be low and is likely to be below the signal-to-background of our experimental diffraction conditions. The forbidden reflections appear in FFT taken from HRTEM images as here the measured interference pattern comes from a highly localized area, which contains a much higher proportion of monolayer material, allowing the weak (110) reflections to be visible in the power spectrum of the HRTEM image.

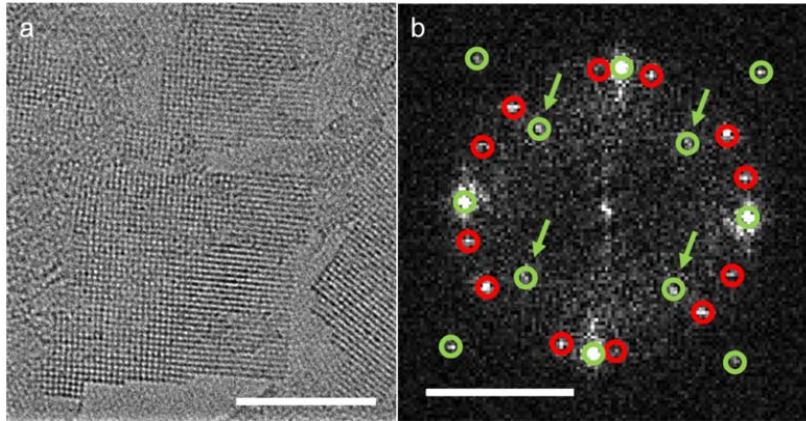

**Figure S9. HRTEM imaging of the thinnest MgO crystallites:** (a) HRTEM image of the thinnest MgO crystallites found encapsulated between two graphene sheets (scale bar 5 nm). (b) Fast Fourier transform (FFT) obtained from the image in Fig. S9a showing lattice reflections from the two graphene sheets (red circles) and from the MgO crystals (green circles). The arrows indicate the presence of a (110) lattice reflection which is usually forbidden for bulk MgO (scale bar 5 nm$^{-1}$).

7. Proposed MgCl$_2$ Reaction Mechanism

Magnesium chloride is one of the typical ionic halides that has various hydrated states MgCl$_2$(H$_2$O)$_x$. Magnesium chloride hexahydrate, MgCl$_2$(H$_2$O)$_6$, is a common hydrate, and it can be simultaneously dehydrated and decomposed at 415 °C to produce MgO[24]. Magnesium salts can also be converted to Mg(OH)$_2$ through a hydrothermal reaction and a subsequent calcination produces MgO[25,26]. Because MgCl$_2$(H$_2$O)$_6$ is soluble in water, the corresponding equilibrium between the solid and liquid phases is as follows:

$$\text{MgCl}_2(6\text{H}_2\text{O}) \rightleftarrows \text{Mg}^{2+}(\text{aq}) + 2\text{Cl}^-(\text{aq}) + 6\text{H}_2\text{O} \quad (1)$$

The ionization of water strongly increases with increasing pressure. $K_w$ (the ionic product or ionization constant of water) increases by two orders of magnitude when the pressure increases from 0.1 MPa to 1 GPa[27]. Therefore, we have an additional reaction in water when the pressure is high, expressed as

$$\text{H}_2\text{O} \rightarrow \text{H}^+ + \text{OH}^- \quad (2)$$

Because MgCl$_2$ is a salt composed of a weak base and a strong acid, when these salts are dissolved in water, they produce acidic solution and undergo hydrolysis[28]. Mg$^{2+}$ ions in the aqueous solution react with the available OH$^-$ ions from the ionization of water to form Mg(OH)$_2$. Consequently, the mechanism that is assumed to lead to the formation of MgO precipitates in our experiment in the presence of high vdW pressure (T = 298K, P ≈ 1.2 GPa) can be separated into two steps:

i)  $\text{Mg}^{2+} + 2(\text{OH}^-) \rightarrow \text{Mg(OH)}_2$            (3)
    $\text{Cl}^- + \text{H}^+ \rightarrow \text{HCl}$                  (4)
ii) $\text{Mg(OH)}_2 \rightarrow \text{MgO} + \text{H}_2\text{O}$            (5)

The first step results in the production of a salt hydroxide and HCl, which is due to the salt hydrolysis reaction. The second reaction is the decomposition of hydroxide to oxide and is due to the confinement and pressure effects caused by the encapsulation (see below). The two steps can be combined into a single reaction mechanism:

$$MgCl_2(6H_2O) \rightarrow MgO+5H_2O+2HCl \qquad (6)$$

8. Other graphene-encapsulated salts and compounds ($Ca(OH)_2$, CsI and $CuSO_4$)

We have also performed encapsulation experiments using aqueous solutions of $Ca(OH)_2$, CsI and $CuSO_4$ and investigated the resulting materials using Raman spectroscopy and STEM analysis. These experiments provided further insight into the mechanism of the vdW pressure induced chemical changes observed above. Fig. S10a compares the Raman spectrum of GE-$Ca(OH)_2$, to bulk $Ca(OH)_2$ and CaO prepared by annealing $Ca(OH)_2$ at 350 °C in vacuum. The spectrum obtained from bulk $Ca(OH)_2$ shows Raman modes at 357, 678, 1087 and 3620 cm$^{-1}$ all of which are assigned to pure $Ca(OH)_2$[29] except the peak 1087 cm$^{-1}$, which appears due to the small amount of inevitable $CaCO_3$ traces[29]. The Raman spectrum obtained from flat regions of the GE-$Ca(OH)_2$ sample show different features compared with bulk $Ca(OH)_2$. Interestingly, it shows similar signatures of the Raman spectrum of CaO with small traces of $Ca(OH)_2$[29]. Apart from the weak traces of $Ca(OH)_2$, the bands at 1090, 1785 and 1933 cm$^{-1}$ in the encapsulated samples are due to the luminescence bands of CaO[29]. All of these observed Raman features in the encapsulated samples are in good agreement with the reference CaO sample as shown in Fig. S10a.

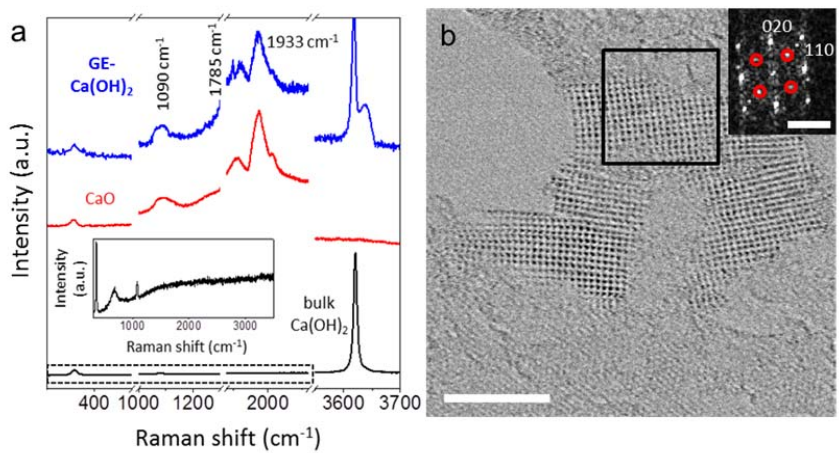

**Figure S10. Graphene-encapsulation-induced conversion of $Ca(OH)_2$ to CaO:** (a) Raman spectrum (633 nm excitation) of bulk $Ca(OH)_2$, reference CaO synthesized by annealing $Ca(OH)_2$ in vacuum at 350 °C and GE-$Ca(OH)_2$. The peak positions of all observed PL bands from CaO are marked. Inset shows the zoomed Raman spectrum of bulk $Ca(OH)_2$ obtained from the marked region. (b) HRSTEM BF image showing the square lattice of CaO in the GE-$Ca(OH)_2$ sample (scale bar 2 nm). Inset shows the FFT from the marked region in the main figure, red circles indicate (110) lattice reflections of CaO (scale bar 5 nm$^{-1}$).

Our TEM analysis of GE-$Ca(OH)_2$ samples further confirms the transformation of $Ca(OH)_2$ to CaO. Fig. S10b shows the HRTEM image of atomically thin CaO crystals formed in the flat GE regions of the $Ca(OH)_2$ samples. The observed lattice constant of CaO ($\approx$ 2.4 Å) is consistent with the reported lattice constant of bulk CaO. Similar to the case of MgO, the forbidden (110) lattice reflections (marked with red circles) are visible in the FFT of thin CaO crystals (inset of Fig. S10b). This implies that the thinnest CaO crystals found in the flat encapsulated regions are monolayers. All the above observations confirm that the decomposition of hydroxide compounds into oxides is possible at room temperature inside nanometer thick graphene enclosures due to the presence of GPa vdW pressure. Similar room-temperature decomposition of hydroxide compounds were reported before in mechanochemistry, where hydroxide compounds were decomposed by grinding with $SiO_2$[30], demonstrating that mechanical force can provide the energy needed to initiate the decomposition as an alternative to conventional thermal energy.

We have also studied GE-CsI and $CuSO_4$ samples using TEM and Raman spectroscopy. Fig. S11a shows the high angle annular dark field (HAADF) STEM image and EELS analysis of a GE-CsI sample. Unlike $MgCl_2$, these measurements are consistent with bulk CsI and we have not found any evidence for chemical reaction in the case of this salt. This can be explained by the absence of a hydrolysis reaction in CsI, as observed in the case of $MgCl_2$. When CsI is dissolved in water, the relative proportions of $H^+$ and $OH^-$ ions in solution remain unchanged, and the solution remains neutral. In other words, the salts of strong acids and bases are completely ionized in water and do not undergo hydrolysis[28]. $CuSO_4$ is an example of a salt of a weak base and a strong acid and this compound undergoes hydrolysis in a similar way to $MgCl_2$. Our encapsulation experiments for $CuSO_4$ indicate that here again chemical changes result due to vdW pressure between the graphene sheets. Fig. S11b shows the Raman spectrum obtained from bulk $CuSO_4$ (pentahydrate) and GE-$CuSO_4$. The spectrum of bulk $CuSO_4$ shows the major Raman bands corresponding to symmetric stretching ($v_1$) vibration at 984 cm$^{-1}$, the symmetric bending ($v_2$) modes between 425–468 cm$^{-1}$, the antisymmetric stretching ($v_3$) modes at 1097 and 1143 cm$^{-1}$, the antisymmetric bending ($v_4$) mode at the 613 cm$^{-1}$ and 248–283 cm$^{-1}$ bands due to the internal modes of the complex. In addition, we have also observed broad bands at 3200–3500 cm$^{-1}$ due to water stretching modes. All of these observed Raman bands are in good agreement with previously reported studies[31,32]. However, the Raman spectra from flat, 1-nm-thick regions of GE-$CuSO_4$ show different Raman bands at 301, 346 and 627 cm$^{-1}$, which are in good agreement with the Raman modes of copper oxide (CuO), namely, 297 cm$^{-1}$ $A_g$ mode, 344 cm$^{-1}$ $B_{1g}$ mode and 629 cm$^{-1}$ $B_{2g}$ mode[33].

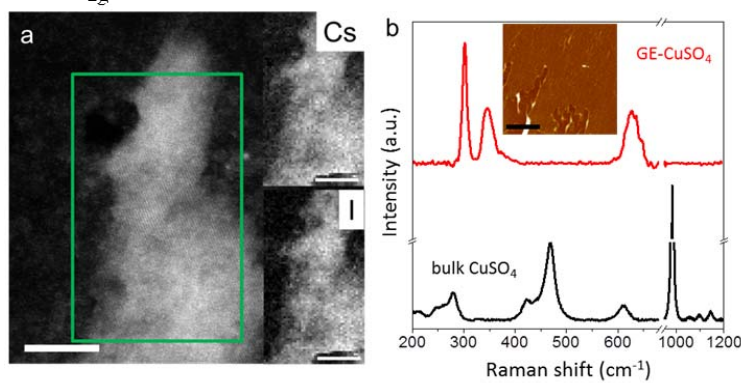

**Figure S11. Graphene-encapsulation effects on CsI and $CuSO_4$ samples:** (a) HAADF STEM image showing a GE-CsI crystal. Inset shows EELS mapping of Cs and I, confirming the co-location of these two elements in the sample (all scale bars are 10 nm). Mapped area is indicated by the green rectangle in the main figure. (b) Raman spectrum (514 nm excitation) of bulk $CuSO_4$ and GE-$CuSO_4$. Inset shows the AFM image of 1-nm-thick, flat, encapsulated regions of $CuSO_4$ (scale bar 1 μm).

9. Stability of nano-enclosed salts

To investigate the stability of enclosed hydrated $MgCl_2$, we performed density functional theory (DFT) calculations using the GAUSSIAN package (G09)[34]. This is an electronic-structure package that uses Gaussian type orbitals as a basis; in this study, the 6-31G* basis set was used. The exchange-correlation (XC) hybrid functional M06-2X was used in G09. The combination of the basis set and XC functional are expected to produce a reliable description of the electronic properties of our system. The self-consistency loop iterates over small changes in the atomic positions until the change in the total energy is less than $10^{-7}$ eV. The geometries are then considered to be relaxed once the force on each nucleus is less than 50 meV/Å. Natural bonding orbital (NBO) analysis[35] was performed with the M06-2X/6-31G* level of theory.

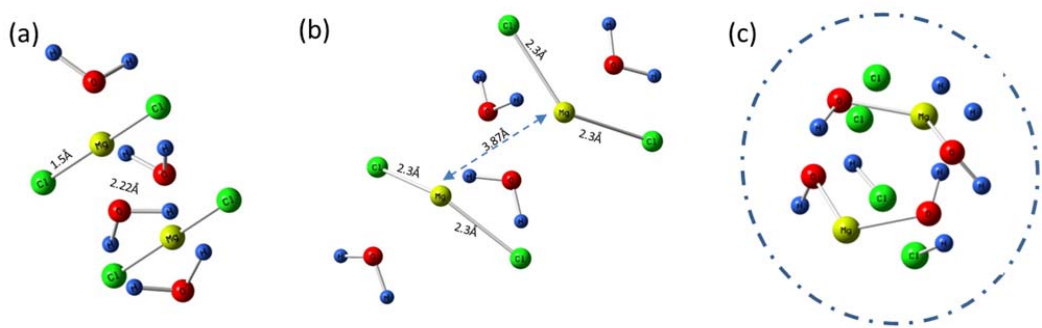

**Figure S12. $MgCl_2$ under confinement.** (a) Initial configuration of compact $MgCl_2$ and four water molecules. (b) Optimized configuration of $MgCl_2$ and water in free space and (c) optimized configuration inside $C_{70}$. The dashed circle in (c) refers to the boundary of $C_{70}$.

Here, we have used a $C_{70}$ fullerene molecule as a hydrophobic confinement system. This is a simple model for studying trapped molecules inside a hydrophobic cage, which is computationally less expensive than a graphene enclosure. The initial configuration of the $MgCl_2$ and water molecules in free space is shown in Fig. S12(a), where four water molecules are randomly placed around two $MgCl_2$. The final optimized configurations in free space and within the confinement are shown in Figs. S12(b) and (c), respectively. In free space, the starting structure for $H_2O$ and $MgCl_2$ was compact, as shown in Fig. S12a. At the end of the structure optimization, the increased separation between $H_2O$ and $MgCl_2$, as shown in Fig. S12b, suggests that the final free space structure is still four $H_2O$ molecules and two $MgCl_2$. Within the confinement, $C_{70}$, $H_2O$ and $MgCl_2$ are all included in the optimization, and the final configuration includes MgO, 2HCl, $2H^+$ and $2Cl^-$ ions, as shown in Fig. S12c. Each O atom in the Mg–O bonds should be saturated by a H atom. These additional –H bonds guarantee the convergence of the DFT calculations because the valance of the oxygen atom is 2. These H atoms would be replaced by other Mg atoms if a larger volume with more Mg atoms was simulated. The Mg–O bond lengths in our DFT optimization are in the range 1.88–1.946 Å, which agrees with the bulk MgO lattice constant (≈ 2 Å). We do not claim that this optimization procedure is equivalent to the precise reaction mechanism; however, it explicitly shows that salts enclosed in extremely small volumes are unstable and behave differently from their bulk counterparts**.** We have also performed additional DFT calculations using confined $Mg(OH)_2$ and three water molecules inside a $C_{70}$. The optimized configuration consists of four Mg–O bonds (terminated with H), one $H_2O$ and one $H_2$. These simulations further show that the MgO is energetically more stable inside extremely small volumes than $MgCl_2$ and $Mg(OH)_2$. Similar structural changes were also observed when we used $C_{90}$ or $C_{180}$ as the molecular enclosure system.

To study the effects of confinement on CsI, we performed similar DFT calculations for two CsI and six water molecules enclosed in $C_{90}$. After optimization, we do not see any signature of Cs–O bond formation; instead, all the oxygen atoms of water prefer to be far from the Cs ion, and the optimized structure includes $2Cs^+$, $2I^-$ ions together with $2OH^-$, $2H^+$ and four water molecules. This further agrees with our experimental observations and supports the proposed reaction mechanism.

10. Defect formation in graphene due to a chemical reaction

Fig. S13 shows the Raman spectrum of graphene obtained from the GE-$MgCl_2$ sample with the presence and absence of MgO in the enclosure. The appearance of a defect-induced D band at 1345 cm$^{-1}$ and a D′ band at 1620 cm$^{-1}$ from the regions where MgO is present in and absence of these bands in the MgO-absent region indicates that the defects are created in the graphene lattice during the chemical conversion of $MgCl_2$ to MgO. Similar features were also observed in the case of GE-$CuSO_4$ samples. We explain this formation of defects in the graphene as an outcome of aggressive chemical reaction that occurs inside the graphene enclosure. As explained in the main text and above, hydrated $MgCl_2$ trapped inside the graphene enclosure experiences high pressure and undergoes a chemical reaction yielding solid MgO along with the vapors of HCl and $H_2O$ as by-products (while in the case of $CuSO_4$, products are CuO, $H_2SO_4$ and $H_2O$). The reactive acids such as HCl and $H_2SO_4$ are known to create defects in graphene and could explain the observed D-peak[36,37].

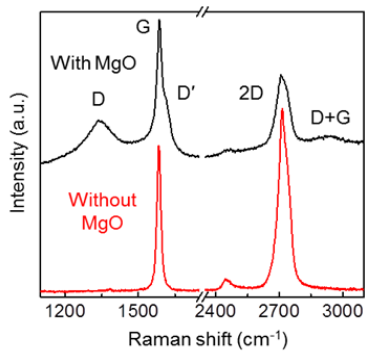

**Figure S13. Defect formation in graphene.** Raman spectrum (488 nm excitation) of graphene (from both the top single layer and bottom few layers) obtained from GE-$MgCl_2$ sample in the presence and absence of MgO, showing the formation of defects in the graphene when MgO is formed in the enclosure.

11. VdW pressure and its effects on other 2D crystal enclosures

To examine the presence of vdW pressure and its effects in other graphene-like 2D crystals, Raman experiments have also been performed on thin BN-encapsulated molecules/salts on few-layer BN. Fig. S14 shows the Raman spectrum of BN-encapsulated BA on few-layer BN. A clear blue shift of 8.8–10 cm$^{-1}$ in the $\nu_3$ ($A_g$) symmetric B–O bond stretching mode and an apparent red shift in the O–H bond stretching modes are observed for ≈ 1-nm-thick flat regions of BN-encapsulated BA. From the observed shift in $\nu_3$ ($A_g$) mode, the estimated vdW pressure in BN nano-enclosures is ≈ 1.9–2.2 GPa, which is slightly higher than that of the graphene nano-enclosures. This increased pressure could be attributed to the strong vdW interactions between BN sheets along with additional charge–charge interactions due to the polar nature of BN. In corroboration with the above experiment, the Raman spectrum of BN-encapsulated $MgCl_2$ has also shown the Raman band at 125 cm$^{-1}$ assigned to MgO, similar to the case of GE-$MgCl_2$ samples.

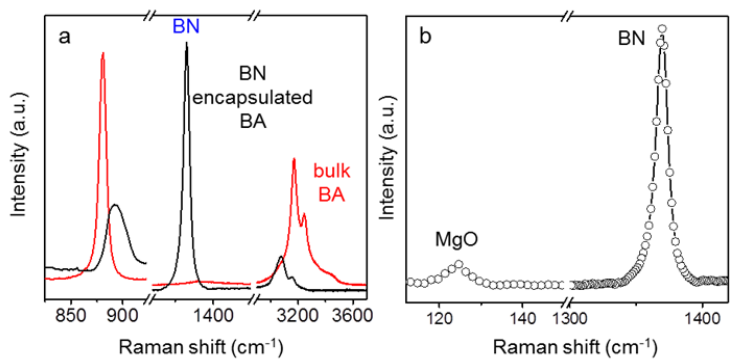

**Figure S14. VdW pressure and its effect on BN encapsulation** (a) Raman spectrum (514 nm excitation) of bulk BA and BN-encapsulated BA on BN showing the vdW pressure-induced Raman shift. (b) Raman spectrum (488 nm excitation) of BN-encapsulated $MgCl_2$ on BN showing the presence of the MgO band and BN band.

As an additional example, we have studied thin mica-encapsulated BA and $MgCl_2$ on mica samples; however, we have not found any signatures of the trapped molecules or salts between the mica sheets. The only signatures we found in our Raman study was the presence of water (O–H band at 3100–3600 cm$^{-1}$) between the mica layers. We attribute this to the high hydrophilicity of mica, which favors the formation of adlayers of water[38,39] and hinders other molecules or salts from being trapped at the interface.